\input harvmac
\overfullrule=0pt

%macros
%
\def\sqr#1#2{{\vbox{\hrule height.#2pt\hbox{\vrule width
.#2pt height#1pt \kern#1pt\vrule width.#2pt}\hrule height.#2pt}}}
\def\Box{\mathchoice\sqr64\sqr64\sqr{4.2}3\sqr33}

\Title{ \vbox{\baselineskip12pt
\hbox{hep-th/0109096}}}
%\hbox{IASSNS-HEP-99/66}
%\hbox{CAL}}}
{\vbox{\centerline{Conformal Operators}
\bigskip
\centerline{for Partially Massless States}}}
\smallskip
\centerline{Louise Dolan
%\foot{dolan@physics.unc.edu}
}
\smallskip
\centerline{\it Department of Physics}
\centerline{\it
University of North Carolina, Chapel Hill, NC 27599-3255}
\bigskip
\smallskip
\centerline{Chiara R. Nappi
%\foot{cnappi@.princeton.edu}
}
\smallskip
\centerline{\it Department of Physics, Jadwin Hall}
\centerline{\it Princeton University, Princeton, NJ 08540}
\bigskip
\smallskip
\centerline{Edward Witten}
%\foot{witten@ias.edu}
\smallskip
\centerline{\it School of Natural Sciences, Institute for Advanced Study}
\centerline{\it Olden Lane, Princeton, NJ 08540, USA}
\bigskip

\noindent
The AdS/CFT correspondence is explored for ``partially massless'' fields
in AdS space (which have fewer helicity states than a massive field but more
than a conventional massless field).  Such fields correspond in the boundary
conformal field theory to fields obeying a certain conformally-invariant
differential equation that has been described by Eastwood {\it et al.} 
The first descendant of such a field is a conformal field of negative norm.
Hence, partially massless fields may make more physical sense in de Sitter
as opposed to Anti de Sitter space. 

\Date{}
%references

\nref\dwone{S. Deser and A. Waldron, ``Null Propagation of Partially
Massless Higher Spins in (A)dS and Cosmological Constant Speculations'',
hep-th/0105181.}

\nref\dwtwo{S. Deser and A. Waldron,
``Partial Masslessness of
Higher Spins in (A)dS'', hep-th/0103198.}

\nref\dnone{S. Deser and R. Nepomechie, ``Gauge Invariance versus
Masslessness in de Sitter Spaces'', Annals of Physics {\bf 154} (1984)
396.}

\nref\dntwo{S. Deser and R. Nepomechie,
``Anomalous Propagation
of Gauge Fields in Conformally Flat spaces'',
Physics Letters {\bf 132B} (1983) 321.}

\nref\dwthree{S. Deser and A. Waldron, ``Stability of Massive
Cosmological Gravitons'', Phys.Lett. {\bf B508} (2001) 347-353,
hep-th/0103255.}

\nref\dwfour{S. Deser and A. Waldron, ``Gauge Invariances
and Phases of Massive Higher Spins in (A)dS'',
Phys.Rev.Lett. {\bf 87} (2001) 031601, hep-th/0102166.}

\nref\hone{A. Higuchi, ``Massive Symmetric Tensor Field in Spacetimes
with a Positive Cosmological Constant'',
Nuclear Physics {\bf B325} (1989) 745.}

\nref\htwo{
A. Higuchi, ``Forbidden Mass Range for Spin-2 Field
Theory in De Sitter Spacetime''
Nuclear Physics {\bf B282} (1987) 397.}

\nref\hthree{A. Higuchi,
``Massive Symmetric Tensor Field in Curved
Spacetime'', Class. Quantum Grav. {\bf 6} (1989) 397.}

\nref\hfour{A. Higuchi,
``Symmetric Tensor Spherical Harmonics on the
N-sphere and their Application to the de Sitter Group $SO(N,1)$'',
J. Math. Phys. {\bf 28} no. 7 (1987) 1553.}

\nref\bgp{I. Buchbinder, D. Gitman and B. Pershin,
``Causality of Massive Spin 2 Field in External Gravity'',
Phys.Lett. {\bf B492} (2000) 161-170,
hep-th/0006144.}

\nref\bgkp{I. Buchbinder, D. Gitman, V. Krykhtin and B. Pershin,
``Equations of Motion for Massive Spin 2 Field Coupled to
Gravity'', Nucl.Phys. {\bf B584} (2000) 615-640,
hep-th/9910188.}

\nref\eastone{M. Eastwood, ``Notes on Conformal Differential
Geometry'',
(available electronically:\break
ftp://ftp.maths.adelaide.edu.au/pure/meastwood/srni95.ps),
Suppl.Rendi.Circ.Mat.\break Palermo {\bf 43} (1996) 57--76.}

\nref\easttwo{R.J. Baston and M. Eastwood, ``Invariant Operators,'' in the
collection Twistors in Mathematics and Physics, L.M.S.
Lecture Notes {\bf 156}, C.U.P. 1990, pp. 129--163.}

\nref\eastthree{M. Eastwood and J.W. Rice, ``Conformally invariant
differential operators on Minkowski space and their curved
analogues.'' Comm.Math.Phys. {\bf 109} (1987) 207--228. Erratum,
Comm.Math.Phys. {\bf 144} (1992) 213.}

\nref\eastfour{M. Eastwood and J. Slovak, ``Semi-holonomic Verma modules,''
Jour.Alg. {\bf 197} (1997) 424--448.}

\nref\witten{E. Witten, ``Quantum Gravity in De Sitter Space'',
hep-th/0106109.}

\nref\strom{A. Strominger, ``The dS/CFT Correspondence'',
hep-th/0106113.}

\nref\gf{C. Fefferman and C.R. Graham, ``Conformal Invariants'',
{\it Elie Cartan et les Math\'matiques d'Aujourdhui} (Asterisque, 1985),
95.}

\nref\hsone{M. Henningson and K. Skenderis, ``The Holographic
Weyl Anomaly'', JHEP {\bf 9807} (1998) 023, hep-th/9806087.}

\nref\hstwo{M. Henningson and K. Skenderis,
``Holography and the Weyl Anomaly,
Fortsch.Phys. {\bf 48} (2000) 125-128, hep-th/9812032.}

\nref\gw{C.R. Graham and E. Witten, ``Conformal Anomaly of Submanifold
Observables in AdS/CFT Correspondence'',
Nucl.Phys. {\bf B546} (1999) 52-64, hep-th/9901021.}

\nref\gkp{S. Gubser, I. Klebanov, and A. Polyakov,
``Gauge Theory Correlators from Non-Critical String Theory'',
Phys.Lett. {\bf B428} (1998) 105-114, hep-th/9802109.}

\nref\w{E. Witten, ``Anti-de Sitter Space and Holography'',
Adv.Theor.Math.Phys. {\bf 2} (1998) 253-291, hep-th/9802150.}

\nref\gub{O. Aharony, S. Gubser, J. Maldacena, H. Ooguri, and
Y. Oz, ``Large N Field Theories, String Theory, and Gravity'',
Phys.Rept. {\bf 323} (2000) 183-386, hep-th/9905111.}

\nref\min{S. Minwalla, ``Restrictions Imposed by Superconformal
Invariance on Quantum Field Theories'',
Adv.Theor.Math.Phys. {\bf 2} (1998) 781-846,
hep-th/9712074 and references therein.}

%\nref\brfone{P. Breitenlohner and D. Freedman, ``Positive Energy
%in Anti-De Sitter Backgrounds and Gauged Extended Supergravity,''
%Phys. Lett. {\bf B115} (1982) 197.}
%
%\nref\brftwo {L. Mezincescu and P.K. Townsend,
%``Stability at a 
%Local Maximum in Higher-Dimensional Anti-De Sitter Space and
%Applications to Supergravity'', Annals of Phys. {\bf 160} (1985)
%406.}

%\nref\brfthree{P. Breitenlohner and D. Freedman, ``Stability in 
%Gauged Extended Supergravity'',  Annals of Phys. {\bf 144} (1982) 249.}
%
%\nref\brffour{P. Townsend, ``Positive Energy and the Scalar Potential
%in Higher-Dimensional (Super)gravity Theories'', Phys. Lett. {\bf B148} 
%(1984) 55.} 

%paper
\newsec{Introduction}

In four-dimensional Minkowski space, a massless field of spin $s$ has
helicities $\pm s$, while a massive field has all possible helicities
$-s,-s+1,\dots,s$.  When the cosmological constant is nonzero, however,
the range of possibilities is greater \refs{\dwone - \bgkp}. 
In addition to the familiar massless
and massive fields, one can have ``partially massless'' fields whose
helicity  ranges over the set $-s,-s+1,\dots,s$ with $-n,-n+1,\dots n$
removed, for any $n\leq s-2$.  The mass squared of such a field is
\eqn\msqd{m^2={\Lambda\over 3} \left(s(s-1)-n(n+1)\right).}
Here we define $\Lambda$ by writing the Einstein equation as 
$R_{\mu\nu}=\Lambda g_{\mu\nu}$.

A partially massless field is described by a symmetric tensor field
$\phi_{\mu_1,\dots,\mu_n}$ with a gauge invariance 
\eqn\harry{\delta\phi_{\mu_1\mu_2\dots
\mu_s}=D_{\mu_1}\dots D_{\mu_{s-n}}\xi_{\mu_{s-n+1}\dots \mu_s}
+\dots,} 
where the $+\dots$ refers to terms obtained by
symmetrizing the indices and adding additional
contributions with fewer than $s-n-1$ derivatives.  
We will describe later in more detail the first nontrivial case,
with $s=2$ and $n=0$.

Formally, partially massless fields can be defined for either positive
or negative cosmological constant.  We note, however, from \msqd\ that
for $\Lambda<0$, a partially massless field has negative mass squared, smaller
than that of  a massless field which has more gauge invariance
-- a result that seems unintuitive.  By contrast, for
$\Lambda>0$, the partially massless field has a positive mass squared.

In this paper, we will study partially massless fields for $\Lambda<0$
using the AdS/CFT correspondence.  Just as a massless field of spin
$s$ corresponds in the boundary theory to a rank $s$ 
conserved symmetric tensor,
a partially massless field must correspond in the boundary theory to a 
partially
conserved tensor.  Such a tensor is a field $L^{\mu_1\dots \mu_s}$, symmetric
in its indices, and obeying a conformally invariant equation whose general
form we can guess given \harry.  
Since a coupling 
$\int L^{\mu_1\dots \mu_s}\phi_{\mu_1\dots \mu_s}$
should be gauge-invariant, the equation must have the form
\eqn\ungu{D_{\mu_1}D_{\mu_2}\dots D_{\mu_{s-n}}L^{\mu_1\dots\mu_s}+\dots = 0}
where the $+\dots$ refers to symmetrization of indices and terms of lower order
that are proportional to the curvature tensor of the boundary.
This equation must be conformally invariant, since the boundary of AdS space
has only a conformal structure, not a metric.  In fact, the appropriate
conformally invariant differential equations have been described in a series
of papers \refs{\eastone - \eastfour}.
We explain below the derivation of \ungu, starting
with a partially massless field in the bulk, for the first nontrivial case
of $s=2,$ $n=0$; the same idea should work in general.

Conformal invariance of \ungu\ holds only if $L$ is assigned a definite
conformal dimension.  This corresponds in the AdS/CFT correspondence to 
the fact that the partially massless field has a definite $m^2$ given in 
\msqd; from that point of view, the conformal dimension of $L$ 
is determined, in a standard
fashion, from the behavior of $\phi$ near the boundary of
AdS space.  

A third way to determine the conformal dimension of $L$ is
as follows.  Via the operator-state correspondence
of conformal field theory, the field $L^{\mu_1,\dots,\mu_s}$
corresponds to a state $|\Psi^{\mu_1,\dots,\mu_s}\rangle$, which
(given that $L$ transforms covariantly under Weyl transformations) 
must be a highest weight vector for the conformal group.  The equation
\ungu\ for $L$ means that a certain level $s-n$ 
descendant of $|\Psi\rangle$ is a null
vector.  This occurs only for a particular conformal dimension for $L$, which
can be conveniently computed using radial 
quantization, and agrees with the results obtained by other methods.  

In sum, then, this paper is devoted to the correspondence 
between three types of objects:

{\it (A)}  A partially massless field $\phi$ in AdS space.

{\it (B)} A field $L$ on the boundary of AdS space that obeys the conformally
invariant equation \ungu. 

{\it (C)} A highest weight vector $|\Psi\rangle$ of the conformal group
in a representation that has a certain null vector at level $s-n$.

Our analysis of {\it (C)}
 also shows that a certain descendant of $|\Psi\rangle$ at a lower
level  has negative norm.  
Despite the nice consistency between {\it (A)}, ({\it B)}, and {\it (C)}, 
this seems discouraging
for most physical applications of partially massless fields in the AdS case.
This may correspond to the strange sign of \msqd\ for $\Lambda<0$.
Alternatively, we could consider partially massless fields in de Sitter
space, with $\Lambda>0$, where \msqd\ has a more intuitive sign.
In this case, many considerations of the AdS/CFT
correspondence can be imitated \witten, \strom\ though the physical 
meaning is less clear.  
Our computations relating {\it (A)}, {\it (B)}, and {\it (C)}
are purely formal, so they make
sense for $\Lambda>0$.  For de Sitter space, the boundary conformal field
theory (if that is the right notion) is anyway not unitary \strom,
so the negative norm descendant is not a problem.

In this paper, we consider only the case of AdS space with bulk dimension
$D=4$ and boundary dimension $d=D-1=3$, 
though the arguments presumably generalize to any
dimension.  In particular, the conformally invariant equations \ungu\ exist
in any dimension \refs{\eastone - \eastfour}. 

In the last section, we present a cosmological 
solution of the boundary conformal field theory
which relates the time dependence of the boundary operator
to the Hubble function.

\newsec{Review of Ingredients}

\subsec{Partially Massless Field Equations}

First, we recall explicitly the field equation
of a partially massless field of spin two and $n=0$, in four dimensions.

We denote as $g_{\mu\nu}$ a background metric that obeys the Einstein equations
\eqn\jundh{R_{\mu\nu}=\Lambda g_{\mu\nu},} and we let $C_{\alpha \mu\nu\rho}$ 
denote the Weyl tensor.  The Riemann and Ricci tensors are defined by  
$R^\rho_{\hskip4pt \mu\nu\lambda} \equiv \partial_\nu 
\Gamma^\rho_{\mu\lambda}
- \partial_\lambda \Gamma^\rho_{\mu\nu}
+ \Gamma^\sigma_{\mu\lambda} \Gamma^\rho_{\sigma\nu}
- \Gamma^\sigma_{\mu\nu} \Gamma^\rho_{\sigma\lambda}\,,$ and
$R_{\mu\nu} \equiv R^\lambda_{\hskip4pt \mu\lambda\nu}\,,$
and $C$ is obtained from the Riemann tensor by subtracting trace terms.

The spin two partially massless field is a symmetric tensor 
$\phi_{\mu\nu}$ whose field equation in $D$ dimensions, 
in a background Einstein spacetime, reads
\eqn\pmeom{(\Box - {{\textstyle{D\Lambda\over (D-1) }}} )
\phi_{\mu\nu}
- D_\mu D_\nu \phi_\rho^\rho
+ {{\textstyle{\Lambda\over (D-1) }}} {g_{\mu\nu}} \phi_\rho^\rho
-2 {C_{\alpha\mu\nu\rho}} \phi^{\alpha\rho} = 0\,.}
 A single divergence of the field equation
gives the constraint
$D^\mu \phi_{\mu\nu} - D_\nu \phi_\rho^\rho = 0.$
The equations of motion have a scalar gauge invariance given by
\eqn\sgi{\phi_{\mu\nu} \rightarrow \phi_{\mu\nu}
+ D_\mu D_\nu \xi + {{\textstyle{\Lambda\over (D-1)}}}
g_{\mu\nu} \xi\,.}
The precise form of \pmeom\ is determined by ensuring this gauge invariance.
In verifying gauge invariance, one makes use of \jundh.

A nonzero $\phi$ field has a nonzero stress tensor, and so physically, in the
presence of such a field, one should no longer impose the vacuum Einstein
equations \jundh; there should be additional
contributions quadratic in $\phi$.  In the presence of such terms, the proof
of gauge invariance of \pmeom\ does not work, so one must add additional
contributions to \pmeom\ and/or \sgi, and possibly a nontrivial transformation
law for the metric, $\delta g\sim \epsilon \phi^2$.  
It has not been demonstrated
in the literature that gauge-invariance of the $\phi$ field can be maintained
exactly
(beyond linear order in $\phi$).  However, we will proceed assuming that
a fully nonlinear theory of partially massless fields does exist.
Assuming that the higher order terms 
can be chosen to maintain gauge invariance, their
details will not concern us.

\subsec{Metric Near The Boundary Of AdS Space}

We take $\Lambda<0$, and consider an Einstein manifold that is asymptotic
near its conformal boundary -- which we take to be at $u=0$ -- to AdS space.
With a suitable choice of coordinates, the metric $g_{\mu\nu}$
of such a manifold can be expressed as
\eqn\met{ds^2 = -{\textstyle{(D-1) \over{\Lambda}}} \,\, u^{-2}
\,( du^2 + f_{ij}(u, x ) dx^i dx^j )}
where the expansion of $f_{ij}(u, x)$ in powers of $u$ is determined
from the Einstein equations, 
and is found \refs{\gf - \gw} to be, to the order that we will need,
\eqn\mettwo{f_{ij}(u, x) = \tilde g_{ij} - \,u^2\,\,(d-2)^{-1}
( \tilde R_{ij} - {1\over 2 (d-1)}\, \tilde R\, \tilde g_{ij} ) 
\,+ \,O(u^3)} when $d = D-1$. The conformal
metric on the boundary ($u=0$) is $\tilde g_{ij}$.
We do not assume that it is conformally flat or conformally Ricci-flat.

\subsec{Partially Conserved Conformal Operators}

As explained in the introduction, the partially massless field $\phi$
will correspond in the boundary to a symmetric tensor field $L^{ij}$
that will obey a conformally invariant ``partial conservation law.''
The requisite conformally invariant equation, which is a special case
of the conformally invariant differential equations studied in \refs{\eastone
- \eastfour}, is in $d$ dimensions
\eqn\pc{D_i D_j L^{ij} + {1\over (d-2)}\, \tilde R_{ij} L^{ij} = 0.}
One can verify directly that this
equation is invariant under the Weyl transformation law
\eqn\weyl
{\eqalign{\tilde g_{ij}(x) &\rightarrow e^{-2\sigma(x)} \,
\tilde g_{ij} (x)\cr
\hskip 5pt L^{ij}(x) &\rightarrow e^{(d+1)\,\sigma(x)}\,
\,L^{ij} (x)\,,}}
since
under \weyl\ 
the Ricci tensor transforms as
$\tilde R_{ij}\rightarrow
\tilde R_{ij} + (d-2) (D_i D_j \sigma + D_i\sigma D_j \sigma)
+ \tilde g_{ij} ( D^k D_k \sigma - (d-2) D_k\sigma D^k \sigma)\,,$
and
$D_i D_j L^{ij} \rightarrow e^{(d+1)\sigma} \,[ D_i D_j L^{ij}
- L^{ij} ( D_i D_j \sigma + D_i\sigma D_j \sigma)\,]\,,$ 
and $L^{ij}$ is traceless.
Note that by successively lowering indices, we get fields $L^i_j$ and
$L_{ij}$ with different Weyl transformation laws, {\it i.e.}
 $L^i_j\to e^{(d-1)\sigma}L^i_j$.  
The weight in the Weyl transformation law of the field
with half its indices up and half down is called the conformal dimension
in conformal field theory. 
So $L$ corresponds to a field of conformal dimension $d-1$, 
and thus of dimension $2$ if $d=3$.

\newsec{Applying The AdS/CFT Correspondence to A Partially Massless Field}

Now we will make the correspondence between {\it (A)} and {\it (B)} as
described in the introduction: we will show that a partially massless spin two
field in the bulk of AdS space corresponds to a conformal field on the boundary
that obeys the partial conservation law.  The basic technique is to study
the behavior of the partially massless field near the boundary of AdS space.
This was done recently in \dwone\ with a different motivation. In this section, we assume $D=4$ and $d=3$.

To solve \pmeom, we impose a gauge condition $\phi^\mu_\mu = 0$,
and use the constraint $D^\mu \phi_{\mu\nu} = 0$
to reduce the equation in $D=4$ to 
\eqn\george{(\Box - {{\textstyle{4\Lambda\over 3 }}} )
\phi_{\mu\nu}
-2 {C_{\alpha\mu\nu\rho}} \phi^{\alpha\rho} = 0\,.}
Following \refs{\dwone, \dwtwo}, we recall the traceless-transverse 
decompositions for vectors and tensors
\eqn\tt{\eqalign{
\phi_\mu^T &= \phi_\mu - D_\mu \,{1\over \Box} \,D\cdot\phi\,,
\qquad D\cdot \phi^T = 0\,,\cr
\phi^{TT}_{\mu\nu} &= \phi_{\mu\nu} 
- D_{(\mu} {2\over {\Box + \Lambda}} (D\cdot\phi_{\nu)})^T
-{\textstyle{1\over 4}} g_{\mu\nu} \phi^\rho_\rho\cr
& \hskip8pt - D_{[\mu} D_{\nu ]} {4\over {\,\Box \,( 3\,\Box \, + 4\Lambda )}}
\,[ D\cdot D\cdot\phi - {\textstyle{1\over 4}} \,\Box \,\phi_\rho^\rho ]\,,\cr
D\cdot &\phi_{\mu\nu}^{TT} = 0 = \phi_\rho^{TT\rho}\,,\cr}}
where $[\ldots ]$ on tensor indices denotes the symmetric traceless part
$M_{[\mu\nu]} \equiv M_{(\mu\nu)} - {\textstyle{1\over 4}} g_{\mu\nu}
M^\rho_\rho$, and $\Box \equiv D^\mu D_\mu$ as usual.
We define the spatially traceless-transverse part $\phi_{ij}\equiv
\phi_{ij}^{TT}$.
The solution \dwone\ that is more singular near $u=0$ is
\eqn\stt{\phi_{ij} (u, x) \sim u^{-1} \phi_{0\,ij} (x)\,,}
so that
\eqn\sttu{\phi_i^{\hskip4pt j} (u, x) =
g^{jk} \phi_{ik} \sim u \,\phi_{0\,i}^{\hskip8pt j} (x)\,.}
In the AdS/CFT correspondence, a bulk field $\phi$ that behaves near the
boundary as   
\eqn\sd{\phi_i^{\hskip4pt j} (u, x)
= u^{d-h} \phi_{0\,i}^{\hskip8ptj} (x)\,}
corresponds \refs{\gkp - \gub}
to a boundary field $L$ of dimension $h$.  So in the present case,
$L$ has conformal dimension two, as expected on the basis of the conformally
invariant partial conservation law \pc.
Note that a second independent solution would replace
\stt\ with $\phi_{ij} (u, x) \sim u^{0} \phi_{0\,ij} (x)$,
which would give a second choice of quantization and result in
the scaling dimension of $1$ instead of $2$.)

By proceeding with more care, we can actually derive the law \pc.
We write $^4D$ and $^3D$ for four- and three-dimensional covariant derivatives
and similarly for the affine connections.
Using the metric \met, we evaluate the gauge transformation law \sgi\
on $\phi_{ij}(u,x)$ near the boundary    as
\eqn\sgiij{\eqalign{
({^4D}_i \,{^4D}_j + {{\textstyle{\Lambda\over 3}}}
g_{ij}) \,\xi
&= (\partial_i\partial_j - {^4\Gamma}^k_{\hskip4pt ij}
\,\partial_k
- {^4\Gamma}^u_{\hskip4pt ij} \,\partial_u
+ \, {{\textstyle{\Lambda\over 3}}}
g_{ij}) \,\xi\cr
&\sim \, - \,\tilde g_{ij}\, ( u^{-2} \xi + u^{-1} \partial_u\xi) \,\cr
& \hskip 14pt + ( {^3D}_i \,{^3D}_j
+ (d-2)^{-1} ( \tilde R_{ij} - {1\over 2 (d-1)}\, \tilde R\, \tilde g_{ij} )\,)
\,\xi\,\cr}}
since to order $O(u^2)$ we find
${^4\Gamma}^k_{\hskip4pt ij} \sim {^3\Gamma}^k_{\hskip4pt ij}$
and ${^4\Gamma}^u_{\hskip4pt ij}\sim u^{-1} \tilde g_{ij}.$
Near the boundary,
\eqn\sttb{\phi_{0\,ij} (x) = u\, {\phi_{ij} (u, x)}}
is independent of $u$ so we take the scalar gauge parameter to behave as
$\xi (u,x) \sim u^{-1} \zeta (x)$.
The leading singularity in \sgiij\ cancels, and
the transformation \sgi\ restricted to the boundary is
\eqn\sgibd{
\phi_{0\,ij} = u\, {\phi_{ij} (u, x)} \rightarrow
\phi_{0\,ij} +  \, ( D_i D_j
+ (d-2)^{-1} ( \tilde R_{ij} - {1\over 2 (d-1)}\, \tilde R\, \tilde g_{ij})\,)
\,\zeta\,.}

In the AdS/CFT equivalence, each field propagating on the AdS space
is paired with an operator in the conformal
field theory.
We consider adding to the Lagrangian a Weyl invariant term
$\int d^3x \,{\sqrt{|\tilde g|}}\, L^{ij} \, \phi_{0\,ij}$
that couples the partially massless field to a conformal operator
$L^{ij}$. Requiring the invariance of this term under the transformation
\sgibd\ and using the tracelessness of $L^{ij}$, we find
after an integration by parts that
\eqn\gipt{
\int d^3x \,{\sqrt{|\tilde g|}}\, ( D_i D_j L^{ij} + (d-2)^{-1}  \tilde R_{ij}
L^{ij} )\, \zeta = 0\,} for all $\zeta$.
This gives \pc\ .

\newsec{Conformal Algebra and Unitary Conformal Field Theory}

Now we turn to study the partially conserved conformal tensor $L$ from the
point of view of conformal field theory and representations of the conformal
algebra.  For this purpose, we take the boundary to be flat and use radial
quantization.
Weyl invariance insures invariance under $H=SO( d+1, 1 )$,
the global Euclidean conformal symmetry group of
$d$-dimensional flat space.
The generators of $H$  are rotations $M_{ij}$, translations $P_i$, 
special conformal transformations $K_i$, and dilatations $D$,
for $1\le i\le d$, with nonzero commutation relations
\eqn\ca{\eqalign{[M_{ij}, M_{rs} ] &=
\delta_{ir} M_{js} - \delta_{is} M_{jr} - \delta_{jr} M_{is}
+ \delta_{js} M_{ir}\cr
[M_{ij}, P_r] &= \delta_{ir} P_j - \delta_{jr} P_i\,,\qquad
[M_{ij}, K_r] = \delta_{ir} K_j - \delta_{jr} K_i\cr
[D, K_i] &= -K_i\,,\qquad
[D, P_i] = P_i\,,\qquad
[K_i, P_j ] = \delta_{ij} D + M_{ij}\,.\cr}}
Using this conformal algebra, we will rederive the conformal dimension of the
field $L$ by requiring that a certain descendant of the state corresponding
to $L$ is a null vector
We will also show that the first descendant state has negative norm.

As reviewed in \min, the Euclidean CFT
radially quantized has a Hamiltonian given by $D$,
and a symmetry algebra $SO( d+1, 1 )$ 
whose generators satisfy hermiticity properties
$M_{ij}^\dagger = - M_{ij}, \,P_i^\dagger = K_i,\,
K_i^\dagger = P_i, \,D^\dagger =  D$. The generators
$M_{ij} = - M_{ji}$ form an $SO(d)$ subalgebra. Conformal operators
are in one to one correspondence with the states of the
conformal field theory,  ${\rm lim}_{x\rightarrow 0} {\cal O}(x)
|0> = |{\cal O}>$. These are eigenstates of the dilatation operator $D$,
with eigenvalue given by the scaling dimension.
The generators $P_i$ ($K_i$) raise (lower) the scaling dimension,
and irreducible representations of the conformal algebra each contain
a primary operator $L(x)$ which satisfies $[K_i, L(0) ]= 0$.
Other states in the representation correspond to descendants
$[P_{i_1},[P_{i_2},\dots[P_{i_k},  L(0)]\dots]]$.

For clarity, along with the partially conserved tensor $L^{ij}$,
we will also consider an ordinary conserved tensor $T^{ij}$ (the usual
example is the stress tensor). $T^{ij}$ obeys an ordinary
conservation law $\partial_i T^{ij}=0$.  This corresponds to vanishing
of the operator 
$[P_i,T^{ij}]$ or equivalently of the  state $P_i|T^{ij}\rangle$.
Similarly, a partially conserved tensor $L^{ij}$ obeys the equation \pc\
which in flat space reduces to $\partial_i\partial_jL^{ij}=0$.
This corresponds to vanishing of the operator $[P_i, [P_j,L^{ij}]]$ or the
state $P_iP_j|L^{ij}\rangle$.  $L$ and $T$ are both traceless and symmetric.
The states $|T^{ij}\rangle $ and $|L^{ij}\rangle$ are primaries,
with
\eqn\huncu{K_s|T^{ij}\rangle=K_s|L^{ij}\rangle=0.}
The transformation law of $|T^{ij}\rangle$ under
rotations is 
\eqn\jucvn{M_{rs}| T_{ij}\rangle = \delta_{ri} |T_{sj}\rangle + \delta_{rj} 
|T_{is}\rangle 
- \delta_{si} |T_{rj} \rangle- \delta_{sj} |T_{ir}\rangle,}
and similarly for $|L^{ij}\rangle.$
To begin with, we denote as $h$ the conformal dimension of $T$ and $L$ and
do not assume that they are conserved or partially conserved.

For the states $P_i|T^{ij}\rangle$ and $P_iP_j|L^{ij}\rangle$
to vanish, their norms must be zero. These norms can be evaluated as follows,
using \huncu. 
For $T$ we have
\eqn\nullone{\eqalign{ || \,|P_i |T^{ij} \rangle ||^2 &=
\langle P_s T^{rs}| P_i \, T ^{ij} \rangle
=\langle T^{rs}|K_sP_iT^{ij}\rangle\cr
&= \langle T^{rs}| \,[ K_s, P_i ] \,T^{ij}  \rangle =
 \langle T^{rs}|( \delta_{s i} D + M_{si} )\,T^{ij}  \rangle\cr
&= (h-d) \langle T^{rs} |  T_s^{\hskip4pt j}  \rangle\,,\cr}}
where we have used \ca\ and \jucvn\
and the fact that $T^{ij}$ is traceless, symmetric and
primary with scaling dimension $h$.  {}From \nullone,
we have that $P_i |T^{ij}>$ is a null state precisely if $h=d$.
This result was to be expected, since $d$ is the standard value of the scaling
dimension of the stress tensor in $d$ spacetime dimensions.
We also see that $P_i|T^{ij}\rangle$ has negative norm if $h<d$.

Evaluating similar commutators, we find
\eqn\nulltwo{\eqalign{ || P_i P_j |L^{ij}\rangle ||^2 &=
\langle L^{rs} |  \,K_r K_s P_i P_j \, L ^{ij} \rangle\cr
&= \langle L^{rs} |  \,K_r \,[ K_s, P_i]\, P_j \, L ^{ij} \rangle
+ \langle L^{rs} |  \,K_r \,P_i \,[ K_s,  P_j ]\, L ^{ij} \rangle\cr
&= 2 \,(h-(d-1))\, \langle L^{rs} | \,K_r  P_j \, L ^{ij} \rangle\cr
&= 2 \,(h-(d-1))\,\, (h - d) \,\,\langle L^{rs} | L_{rs} \rangle\,.\cr}}
So for 
$P_i P_j |L^{ij}\rangle$ to be  a null vector, we require $h=d-1$,
as found earlier.

Applying the same technique, we can show that the first descendant of $L$ is
actually a state of negative norm.
In fact, 
\eqn\nullthree{
||\, |P_i L^{ij}> ||^2
= - < L^{js} |  L_s^{\hskip4pt j} >\,.}
The computation here is  precisely the same as the one in \nullone, so the
result directly reflects the fact that the conformal dimension of $L$ is less
than that of $T$.  In terms of partially massless fields in AdS space,
this corresponds directly to the sign of the formulas for $m^2$ in \msqd.
Thus, as mentioned in the introduction, the physical properties of partially
massless fields in de Sitter space may be better.

What property of the partially massless field $\phi$ in
AdS space corresponds to the
negative norm of the first descendant of $L$ on the boundary?  
The descendants of $L$
all have real conformal dimensions, greater than that of $L$, but one of them
has negative norm.  This seems to mean that $\phi$
 can be quantized with energies bounded below -- and with the Fock
ground state having the smallest energy -- but not in a Hilbert space with all
states having positive norm.  This is a delicate result that will depend upon
analysis of global boundary conditions in AdS space.

Though we have restricted ourselves to fields of spin 2 up to this point,
the computations in this section are readily generalized to an $s^{th}$
rank symmetric traceless tensor $L^{i_1i_2\dots i_s}$.
Looking for null vectors that would correspond to a partial conservation law
\eqn\kij{\partial_{i_1}\dots \partial_{i_m}L^{i_1i_2\dots i_s}=0,} we get 
\eqn\nullfour{
{\eqalign{ &||\, |P_{i_1} P_{i_2} \ldots
P_{i_m}  L^{i_1 i_2\ldots i_s} \rangle ||^2\cr
&= \langle L^{j_1 j_2\ldots j_s}|  \,K_{j_1} K_{j_2} \ldots
K_{j_m} \, P_{i_1} P_{i_2} \ldots P_{i_m}\,
|L ^{i_1 i_2\ldots i_s} \rangle\cr
&= m!  \, (h - (d + s- 2) ) \, (h - (d + s - 3) )
\ldots (h - (d + s - m - 1) ) \,
||\, |L^{i_1 i_2\ldots i_s}  > ||^2.
 }}}
If we want \nullfour\ to vanish for $m=r$ but to be nonvanishing for $m<r$,
so that every partial conservation law obeyed by $L$ is a consequence of \kij,
we must take $h=d+s-r-1$.
Such an $L$ should correspond in AdS space to a partially massless field
with a range of helicities missing, depending on $r$.  The first descendant
always has negative norm, if $r>1$.

%1. For J^a, with P_a J^a = 0, then h = d-1
%2. For T^{ab}, with P_a T^{ab} = 0, then h=d
%   For L^{ab}, with P_a P_b L^{ab} = 0 (but P_a L^{ab}\ne 0), then h = d-1.
%3. For T^{abc}, with P_a T^{abc} = 0, then h = d+1
%   For L^{abc}, with P_a P_b L^{abc} = 0 (P_a L^{abc}\ne 0), then h = d
%   For J^{abc}, with P_a P_b P_c J^{abc} = 0 (but P_a J^{abc}\ne 0,
%and P_a P_b J^{abc}\ne 0), then h = d-1.

%So for spin three, L^{abc} and J^{abc} both will have problems
%with neg norm for descendant states. (note for spin 3 there are
%two distinct partially massless states).

%4. For each spin n \ge 1, there is one massless state with h = d-2+n;
%and n-1 distinct partially massless states with scaling dimensions
%h= d-3+n, d-4+n, ..., d-1.
\newsec{A Cosmological Solution}

Since we do not have a full knowledge of the elementary particles,
it might be that in addition to the usual massless particles such as photons
whose interactions are trivial in the infrared, there are additional
massless modes with infrared-nontrivial interactions, governed by
a non-trivial conformal field theory.  If we neglect the problem
with the negative norm descendant (as we will in the present section),
then we can imagine that this theory might admit a partially conserved
$L$ tensor such as we have studied above.
If so, to characterize the physical state of the expanding universe,
in addition to the usually almost conserved quantities such as baryon
number, energy, and entropy, one must also specialize the value of the
$L$ tensor.

To get an idea of cosmology with the $L$ tensor, we will solve
the partial conservation equation \pc\ in the background of
a $d$-dimensional Robertson-Walker metric with flat spatial sections:
\eqn\rw{\tilde g_{ij}\, dx^i dx^j = dt^2 - f^2(t)\,\delta_{mn}\,  dx^m dx^n.}
We assume that the $L$ tensor is invariant under the spatial rotation
and translation symmetries of the background solution.  This implies
that $L^{00}$ is a function of time only, that $L^{0m}=0$, and
(as $L$ is traceless) that $L^{mn}=\delta^{mn}(d-1)^{-1}f^{-2}L^{00}$.
Solving the partial conservation equation, 
we will express $L^{00}$ in terms of  the conformal time
$\eta = \int^t dt' f^{-1}(t')$. For the metric \rw,
the nonvanishing components of the Ricci tensor are
$\tilde R_{00} = - (d-1)  f^{-1} \partial^2_0\,  f $
and $\tilde R_{mn} =  \delta_{mn} ( (d-2) ( \,\partial_0 f )^2 
+ f \partial_0^2 f \,)\,.$ Then $\tilde R_{ij} L^{ij} 
= - (d-2) (\partial_0 H)
L^{00}$ where $H$ is the Hubble function $H(t)\equiv f^{-1}\partial_0
f\,,$ and the partial conservation law \pc\ becomes
\eqn\cosm{\partial_0^2 L^{00} + (2d-1)\, H \,\partial_0 L^{00}
+ (d-1) (\partial_0 H ) \,L^{00} + d(d-1) H^2 \, L^{00} = 0\,.}
We can factor this as
\eqn\frank{( \partial_0 \, + d\, H\,) \, (\partial_0 \, + (d-1)\, H\,)
\, L^{00} = 0\,,}
and solve it by defining $G(t) \equiv \partial_0 L^{00} \, 
+ (d-1)\, H\, L^{00}$, where $ \partial_0 G = - d H G\,.$
Then either $G= 0\,,$ or $G = \,e^{ -d \int^ t dt' H(t')} = f^{-d}\,.$
For  $G= 0$, we have $L^{00} = e^{-(d-1)\int^ t dt' H(t')}
= f^{-(d-1)}$. For the second solution, when $G = f^{-d}$ then
$L^{00}$ can be expressed as $f^{-(d-1)}\,\eta$ where
$d\eta \equiv dt f^{-1}(t)$
\eqn\conftime
{L^{00}(t) = (f(t))^{-(d-1)} \,\int^t dt' f^{-1}(t')\,.} 
The unique general solution of \pc\ in terms of two arbitrary 
constants $a_1, a_2$ 
for the partially conserved
rotationally symmetric boundary operator is
\eqn\gensol{L^{00} = f^{-(d-1)}\, ( a_1 \, + a_2 \,\eta \,)\,.}

In the spirit of the partial conservation law \pc, we can 
identify the partially conserved charge
\eqn\pccharge{\eqalign{ Q &\equiv \int d^{d-1}x {\sqrt{\tilde g}}\,
D_i \,L^{i0} = \int d^{d-1}x \, f^{(d-1)}\, 
(\partial_0 L^{00} + d\, H\, L^{00})\cr
& = (\int d^{d-1}x \,)\,\, 
(a_1 H + a_2  ( H \,\eta + f^{-1} )\,)\,,\cr}}
where $\partial_0 Q \ne 0$ but 
\eqn\pcctwo{\partial_0 Q = -(d-2)^{-1}\,\int d^{d-1}x {\sqrt{\tilde g}}
\,\tilde R_{ij} \,L^{ij} = \, (\int d^{d-1}x )\,
\, (a_ 1 + a_2 \,\eta\,)\,\partial_0 H \,,}
which is an identity on shell 
since $H \partial_0 \eta + \partial_0 f^{-1} = 0$.

In standard cosmology, the energy-momentum tensor
$T^i_j = {\rm diag\,} ( \rho, -p,-p,-p)$ provides
the evolution of the cosmic scale factor $f(t)$ from the
Einstein field equations as 
\eqn\cscale{f(t) \sim  t^{2/(3(1+\omega))}} 
where $\rho \sim f^{-3\,(1+\omega)}$ and $\rho = \omega p$,
for $\omega$ independent of time.  The standard examples
include photons $f \sim t^{\scriptstyle{1\over 2}}$ for 
$\omega = {\textstyle{1\over 3}}\,;$ 
matter $f \sim t^{\scriptstyle{2\over 3}}$ with $\omega = 0\,;$
de Sitter inflation $f \sim e^t $ for $\omega = -1\,;$
and a curvature dominated model $f \sim t$ with
$\omega = -{\textstyle{1\over 3}}\,.$ 
The explicit form of the cosmological scale factor
enables one to study the rate of speeding up of the expansion: 
the acceleration parameter $q(t)\equiv (\partial_0 f )^{-2}
\partial_0^2 f \, f $ is negative for the first two examples,
positive for the de Sitter universe, and zero for the
last example. 

Thus, we have shown that in the context of cosmology, the partial
conservation law determines the time-dependence of the $L$-field,
which evolves according to a quasi-conservation law somewhat
analogous to that of more familiar almost conserved quantities in
cosmology.

\vskip50pt
{\bf Acknowledgements:}
All authors thank the Caltech-USC Center for Theoretical Physics
where this work was initiated. 
CRN was partially supported by the  U.S. Department of Energy,
Grant No. DE-FG03-84ER40168.
Research of EW was supported in part by
NSF Grant PHY-0070928 and the Caltech Discovery Fund.
LD also thanks the 
Institute for Advanced Study at Princeton for its hospitality,
and was partially supported by the  U.S. Department of Energy,
Grant No. DE-FG02-97ER-41036/Task A.

\listrefs

\bye